\documentclass[twocolumn,a4,10pt]{article}
\usepackage{iasted}
\usepackage{times}
\usepackage[dvips]{graphicx}
\usepackage{bm,amsfonts,amssymb,amsmath}
\usepackage{graphicx}

\renewcommand{\min}{\land} \renewcommand{\max}{\lor}

\begin{document}

\date{}

\title{To which extend is the ``neural code'' a metric ?}

\author{Bruno Cessac, Horacio Rostro-Gonz\'alez, Juan-Carlos Vasquez, Thierry Vi\'eville \\ Laboratoire Jean Alexandre Dieudonn\'e, Nice \& INRIA Cortex \& Odyssee {\footnotesize \tt \em http://inria.fr}}

\maketitle

\thispagestyle{empty}

\noindent
{\bf\normalsize ABSTRACT}

Here is proposed a review of the different choices to structure spike trains, using deterministic metrics. 
Temporal constraints observed in biological or computational spike trains are first taken into account 
The relation with existing neural codes (rate coding, rank coding, phase coding, ..) is then discussed.

To which extend the ``neural code'' contained in spike trains is related to a metric appears to be a key point,
a generalization of the Victor-Purpura metric family being proposed for temporal constrained causal spike trains.

\vspace{2ex}
   
\noindent
{\bf\normalsize KEY WORDS}

Spiking network. Neural code. Gibbs distribution. 

\section{Introduction: spike trains in the real life}

The output of a neural network is a set of events, defined by their occurrence times, up to some precision:
\centerline{${\cal F} =  \{ \cdots t_i^n \cdots \}$, $t_i^1 < t_i^2 < \cdots < t_i^n < \cdots$,} 
where $t_i^n$ corresponds to the $n$th spike time of the neuron of index $i$. 
Such {\em spike train} writes $\rho_i(t) = \sum_{t_i^n \in {\cal F}_i} \delta(t - t_i^n)$ with related inter-spike intervals $d_i^{n} = t_i^{n} - t_i^{n-1}$, 
using the Dirac symbol $\delta(.)$. See e.g. \cite{dayan-abbott:01,gerstner-kistler:02b,schrauwen:07} for an introduction.

In computational or biological contexts, not all multi-time sequences correspond to spike trains since it is constrained by the neural dynamic, 
while temporal constraints are to be taken into account: 
Spike-times are:
\\ - [C1] bounded by a refractory period $r, r < d_i^{n+1}$,
\\ - [C2] defined up to some absolute precision $\delta t$, while
\\ - [C3] there is always a minimal delay $dt$ for one spike to be able to interact to another, and
\\ - [C4] %in some context, 
	there is a maximal inter-spike interval $D$ such that either $d_i^{n+1} < D$ or $t_i^{n+1} = +\infty$ (i.e. either neuron fires within a time delay $< D$ or it remains quiescent forever).

For biological neurons, typically, in milliseconds: 
~\\
\centerline{\scriptsize \begin{tabular}{|c|c|c|c|} \hline $r$ & $\delta t$ & $d t$ & $D$ \\ \hline $1-2$ & $0.1-1$ & $> 0.01$ & $10^{2-3}$ \\ \hline \end{tabular}}

The [C1] constraint is well-known, limiting the maximal rate. [C2] seems obvious but is often ``forgotten'' in model. In rank coding schemes for instances \cite{gautrais-thorpe:98} it is claimed that ``all'' spike time permutations are significant, which is not realistic since many of these permutations are indistinguishable, because of the bounded precision, as discussed in e.g. \cite{vieville-crahay:04}. Similarly, a few concepts related to ``liquid states'' \cite{maass:97} assume implicitly an unrealistic unbounded time precision. Similarly, [C3] is also obvious and allows to avoid any causal paradox (e.g.: avalanche effect), but the induced simplifications are not always made explicit. 

The [C4] constraint is less obvious. The idea is that all neurons have a ``leak''. Thus, in the absence of input, the potential decreases and 
neuron cannot  fire anymore. In the brain, current observations also show that a neuron is either firing or . . dead \cite{dayan-abbott:01}. Now [C4] is easily violated for neural model with constant current input, able to integrate during a unbounded period of time, or with maintained sub-threshold oscillations. As discussed in details in \cite{cessac-vieville:08} the fact [C4] is verified or not, completely changes the nature of the dynamics. In the latter case, ``ghost orbit'' occurs: A ``vicious'' neuron can remain silent a very long period of time, and then suddenly fire inducing a complete change in the non-linear system. We distinguish situations with and without [C4] in the sequel.

Considering C[1-3] and eventually [C4], let us now review the related consequences regarding modeling\footnote{They also induce important consequences at the simulation level \cite{cessac-vieville:08b}.}.

\section{The maximal amount of information}

Considering [C1-2], given a network of spiking neurons observed during a finite period $[0, D]$, the number of possible spikes is obviously limited by the refractory period $r$. 
Furthermore, the information contained in all spike times is strictly bounded, since two spike occurrences in a $\delta \tau$ window are not distinguishable.

A rather simple reasoning \cite{cessac-vieville:08b} yields an {\em upper bound for the amount of information}:
\\ \centerline{$N \, \frac{D}{r} \, \log_2\left(\frac{D}{\delta \tau}\right)$ bits during $D$ seconds}
Taking the numerical values into account it means for large $D$, in milliseconds, about $D \, \log_2(D)$ bits/neuron.

In the particular case of fast-brain mechanisms where only ``the first spike matters'' \cite{thorpe-fabre-thorpe:01}, this amount of information is not related to the 
permutations between neuron spikes, i.e. of order of $o(\log(N!)) = N \, \log(N)$  but simply proportional to $N$, in coherence to what is found in \cite{vieville-crahay:04}.

This bound is coherent with results presented in \cite{rieke-et-al:96} considering spike rate and using an information entropy measure. For instance, considering
a timing precision of $0.1 - 1 ms$ as derived here, the authors obtain an information rate bounded around $500 bits/s$ for a neural receptor. 

Note that this is not bad, but good news. For instance in statistical learning, this corresponds to a coding with large margins, thus as robust as support-vector machines,
explaining the surprisingly impressive performances of fast-brain categorization \cite{vieville-crahay:04}.

\section{Dynamics of time-constrained networks}

A step further, taking [C1-3] into account, allows to ``discretize'' the spike trains sequences: i.e. use ``raster\footnote{\label{raster} Formally, 
the spike train discretized raster, writes for $k \geq 0$:
\centerline{$w_i[k] = \#\{ t_i^n, k \, \Delta T \leq t_i^n < (k + 1) \, \Delta T \} \in \{0, 1\}$}.}''. 
The sampling period $\Delta T$ is taken smaller than $r$, $\delta t$ and $d t$. 

In simple models such as basic leaky integrate and fire neuron 
 or integrate and fire neuron models with conductance synapses and constant input current,
this discretization allows a full characterization of dynamics. Thus, it has been shown 
in these two cases that \cite{cessac:08,cessac-vieville:08}:

%This paradigm change yields two major results. For:
%\\ - the B.M.S. model (a minimal discretized integrate and fire neuron model without input), except for a negligible set of initial conditions \cite{cessac:08} and 
%\\ - the gIF model (a discretized integrate and fire neuron model with conductance synapses and constant input current), unless asymptotic trajectories touch thresholds \cite{cessac-vieville:08} 
%(i.e. assuming [C4]) it has been demonstrated that because of the contracting dynamics and the integrate and fire models ``reset'' mechanism:

\begin{itemize}
\item {[H1] \em The raster plot is generically periodic, but, depending on parameters such as external current or synaptic weights,
periods can be larger than any accessible computational time;}
% the increasing period $T$} \\ .\hspace{0.5cm} {\em corresponding to dynamics complexity increase.}
\item {[H2] \em %In the asymptotic stage, 
There is a  one-to-one correspondence between orbits and raster (i.e. raster plots provides a symbolic coding for dynamics).}
\end{itemize}

\begin{figure}[htb]  
 \begin{center} \includegraphics[width=7cm,height=3.5cm]{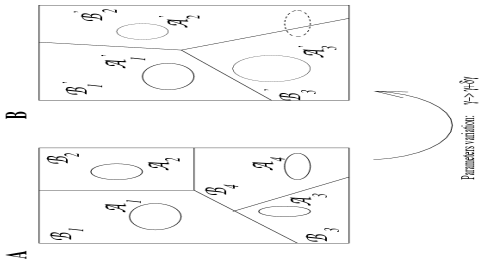} \end{center} 
 \caption{\small \label{attractors} Describing the dynamic landscape of deterministic time-constrained networks.
[{\bf A}] The phase space is partitioned into bounded domains ${\cal B}_l$ and for each initial condition in ${\cal B}_l$ the initial trajectory is attracted to an
attractor, here not a fixed point, as in, e.g. Hopfield networks, but a periodic orbit ${\cal A}_l$.
[{\bf B}] If the parameters (input, weights) change, the landscape is modified and several phenomena can occur: 
change in the basins shape, number of attractors, modification of the attractor as for ${\cal A}_3$ in this example;
A point belonging to ${\cal A}_4$ in Fig.\ref{attractors} A, can, after modification of the parameters, converge
 either to attractor ${\cal A}'_2$ or ${\cal A}'_3$.}
\end{figure}

The fact [H1] allows to clearly understand to which extends spike trains can code information: Periodic orbits give the code. The fact [H2] means that, in these cases, 
the raster is a ``symbolic coding'' in the sense that no information is lost by considering spike times and not the whole neural state. 
Both facts also allow one to deeply understand the network dynamics: Fig.~\ref{attractors} sketches out some aspects.

\vspace{-0.6cm}

To which extends such ``canonical situation'' extends to more complex models is an open question and we can easily conjecture that it is not true for, e.g. 
Hodgkin-Huxley \cite{hodgkin-huxley:52} neuron's assembly. However it is at least true for a large class of computational models actually at the state of the art, 
enlightening the kind of code they may generate.

\section{Neural coding and temporal constraints}

 Let us now introduce the central idea of this review. 

 As an illustrative example, let us consider the temporal order coding scheme \cite{gautrais-thorpe:98,thorpe-fabre-thorpe:01} (i.e. rank coding): only the order of the events matters, not their absolute time values. This means that two spike trains ${\cal F}_1$, ${\cal F}_2$ with the same event ordering correspond to the same code. The key point here, is that rank coding defines a partition of 
spike trains set, each spike train with the same ordering being in the same equivalent class of the partition, i.e. corresponding to the same ``code''. Choosing this code means we have structured spike trains using an ``equivalent relation'' (i.e. ${\cal F}_1$ and ${\cal F}_2$ are equivalent if they correspond to the same code).

 The same view can be given for other coding: rate coding means that all spike trains with the same frequencies are in the same equivalence class, irrespective of their phase, etc.. 

 Let us now introduce a 
%partition metric\footnote{In the topological sense, a partition metric is the discrete metric in the equivalence relation quotient space.}
``distance''
% for any coding and defines this distance 
$d(.)$, which is $0$ if ${\cal F}_1$ and ${\cal F}_2$ correspond to the same code, and $1$ otherwise. 
%This is a true distance function, which completely defines the ``code''.

 The fact that spikes precision is not unbounded leads to many indistinguishable orderings. This does not change the rank coding concept, while the partition is now coarser.
 Trains with two spikes with indistinguishable occurrence times are in the same equivalence class. 

 A step further, how can we capture the fact that, e.g. for rank coding, two spike times with a difference ``about'' $\delta t$ are ``almost'' indistinguishable ? The natural idea is to use, not discrete distances (i.e. with binary 0/1 values) but a 
``quantitative'' distance. 
%A metric. 
Two spike trains correspond exactly to the same neural code if the distance is zero
and the distance increases with the difference between the codes.. 
%To almost  the same code if the distance is, say, below 1. And to different, when not very different codes, if the distance is higher.

 This is the idea we wanted to highlight here. This proposal is not a mathematical ``axiomatic'', but a simple {\em modeling choice}. It is far for being new, but surprisingly enough not made explicit at this level of simplicity. In order to see the interest of this idea, let us briefly review the two main classes of spike train metrics.

 As reviewed in details in \cite{schrauwen:07,victor:05} spike trains deterministic metrics can be grouped in three classes:
\\ -0- ``Binned'' metric, with spikes grouping in bins (e.g. rate coding metrics), not discussed here.
\\ -I-  Convolution metrics, with a distance defined on some convolution of spike train, including raster-plot metric.
\\ -II- Spike time metrics, such as alignment distances \cite{victor-purpura:96}

\section{Using convolution metrics} 

\paragraph{Linear representation.} A large class of metrics derives from the choice of a convolution kernel $K$ writing: 
   \centerline{$s_i(t) = \sum_{t_i^n \in {\cal F}_i}  K_i(t - t_i^n) = K_i * \rho_i \in ]0, 1]$,}
easily normalized between $0$ (no spike) and, say, $1$ (burst mode at the maximal frequency). 
The distance is then defined on the signal ${\bf s} = (\cdots, s_i, \cdots) \in {\cal R}^N$, e.g. using $L^p$ norms. 
The ``code'' here corresponds to the linear representation metric. It allows to link spike trains with a quantitative signal ${\bf s}$.

\begin{figure}[htb]  
 \begin{center} {\small \begin{tabular}{lll}
{[A]}   & \parbox{2cm}{\includegraphics[width=1.5cm]{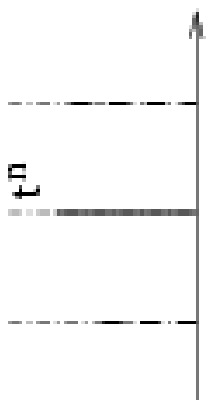}}  & $\delta(t - t_i^n)$ \\
{[B]}  & \parbox{2cm}{\includegraphics[width=1.5cm]{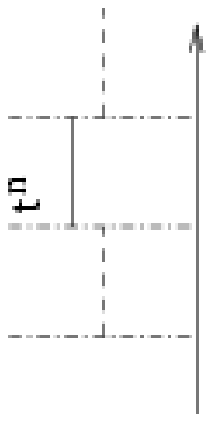}}  & $\chi_{[t_i^{n-1}..t_i^n[}(t) \, \frac{d_{min}}{t_i^n - t_i^{n-1}}$ \\
{[C]} & \parbox{2cm}{\includegraphics[width=1.5cm]{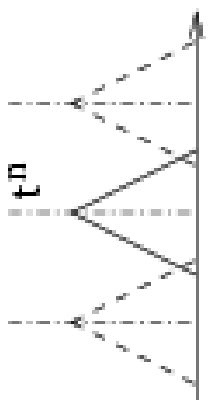}}  & $\mbox{max}\left(0, \frac{d_{min} - |t - t_i^n|}{d_{min}}\right)$ \\
{[D]}  & \parbox{2cm}{\includegraphics[width=1.51cm]{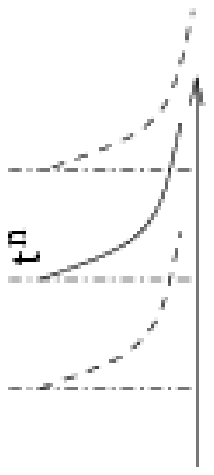}} & $(1-e^\frac{-d_{min}}{\tau}) \chi_{[0..\infty[}(t - t_i^n) \, e^\frac{-(t - t_i^n)}{\tau}$ \\
%factor(expand(sum(exp(-e/t*k),k=0..infinity)*(1-exp(-e/t))));
\end{tabular}} \end{center}
\caption{\small \label{kernel} A few examples of spike train convolution: [A] The spike train itself, [B] A causal local frequency measure estimation (writing $\chi$ the indicatrix function), [C] A non-causal spike density, uniformly equal to 1 in burst mode, [D] A normalized causal exponential profile. Evoked post-synaptic potential profiles model are nothing but such causal convolution (using e.g. double-exponential kernels to capture the synaptic time-constant (weak delay) and potential decay). Similarly spike-trains representations using Fourier or Wavelet Transforms are intrinsically related to such convolutions.}
\end{figure}

\vspace{-0.6cm}

The so-called ``kernel methods'' based on the Mercer theorem \cite{schrauwen:07} are in direct links with linear representation since they are defined, as scalar products, writing:
~\\
\centerline{$k({\cal F}, {\cal F}') = \sum_i \sum_{n, m} \hat{K}_i(t_i^n - t_i^{'m}) = \int_t s_i(t) \, s'_i(t)$}
with direct correspondences for usual kernels with linear convolutions, e.g.:
~\\
 \centerline{\scriptsize \begin{tabular}{|l|c|c|c|} \hline & Triangular   & Exponential  & Gaussian 
\\ \hline
 $K_i(t)$ & $\sqrt{\frac{\lambda}{2}}               \, H\left(t \, \left(\frac{2}{\lambda} - t\right)\right)$
          & $\sqrt{2 \, \lambda}                    \, H(t) \, e^{-\lambda \, t}$
          & $\sqrt{\frac{2 \, \lambda}{\sqrt{\pi}}}         \, e^{-2 \, \lambda^2 \, t^2}$
\\ \hline
 $\hat{K}_i(d)$ & $\mbox{max}\left(1 - \frac{\lambda}{2} \, |d|, 0 \right)$ 
                & $e^{-\lambda \, |d|}$ 
                & $e^{-\lambda^2 \, d^2}$ 
\\ \hline
\end{tabular}}

This also includes distances based on inter-spike intervals as developed in e.g. \cite{kreuz-haas:07}.

Non static kernels of the form $K_i(t, t - t_i^n)$ or $K_i^n(t - t_i^n)$ can also be used (clock-dependent coding, raster plot, 1st spike coding, ..),
while non-linear Volterra series allows to represent ``higher order'' phenomena (see e.g. \cite{rieke-et-al:96}). 
Weighted spike trains (i.e. with quantitative values attached to each spike) are also direct generalizations of these.

These linear representations not only provide tools to compare different spike trains, but allows one to better understand the link between quantitative signals and spike times.
For instance \cite{dayan-abbott:01,maass:97}, writing $s(t) = \sum_i \lambda_i s_i(t)$ allows to define some network readout to link spiking networks to ``analog'' sensory-motor tasks.
Let us illustrate this aspect by the following two results. 

\paragraph{Kernel identification.} Given a causal signal $\bar{\bf s}$ generated by spike train ${\cal F}$ the problem of identifying the related kernel 
is formally solved by the following program:
~\\
\centerline{$\mbox{min}_{K} \int_{t>0} |{\bf s}(t) - \bar{\bf s}(t)|^2 \equiv \int_\lambda |K(\lambda) \, {\bf \rho}(\lambda) - \bar{\bf s}(\lambda)|^2$}
using the Laplace transform Parseval theorem, thus:
~\\
\centerline{$K(\lambda) = \left[ \bar{\bf s}(\lambda) \, \rho(\lambda)^T \right] \, \left[ \rho(\lambda) \, \rho(\lambda)^T \right]^{-1}$}
i.e. the spike train cross-correlation / auto-correlation ratio. Non-causal estimation would consider the Fourier transform. 
This setting corresponds to several identification methods \cite{dayan-abbott:01,schrauwen:07}.

\paragraph{Signal reconstruction.} In order to further understand the power of representation of spike trains in this case \cite{lazar:05} has generalized the well-known Shanon 
theorem, as follows: A frequency range $[-\Omega, \Omega]$ signal is entirely defined by irregular sampling values $s_i^n$ at ``spike time'' $t_i^n$ 
~\\
\centerline{$s_i(t) = \sum_n K_i^n(t - t_i^n)$ {\scriptsize with $K_i^n(t) = s_i^n \, \frac{\sin(\Omega t)}{\pi \, t}$}}
providing that $max_n d_i^n \leq \frac{\pi}{\Omega}$. Thus providing an explicit signal ``decoding''.

\paragraph{Raster-plot metrics.} A step further, it is easy to see, that representing the spike time by a ``raster$^{\mbox{\scriptsize \ref{raster}}}$'' corresponds to non-static convolution kernel. Spike trains can be represented as the real number in $[0 .. 1[$ which binary representation corresponds to the spike-train, inducing new metrics. A useful related metric is of the form, for $\theta \in ]0, 1[$:
~\\
\centerline{$d_\theta({\bf \omega}, {\bf \omega}') = \theta^T, T = \mbox{argmax}_t \;  {\bf \omega}^t = {\bf \omega}'^t$}
thus capturing the fact that two rasters are equal up to a certain rank. Such metrics are used to analyze the dynamics of spiking networks 
and are typically used in the context of symbolic coding in dynamical systems theory\cite{cessac:08,cessac-vieville:08}.

\section{Using alignment metrics}

\paragraph{The original alignment metric.}
The second family of metrics we want to review directly considers spike times \cite{victor-purpura:96,victor:05}. 

The distance between two finite spike trains ${\cal F}$, ${\cal F}'$ is defined in terms of the minimum cost of transforming one spike train into the other.
Two kinds of operations are defined:
\\ - spike insertion or spike deletion, the cost of each operation being set to $1$
\\ - spike shift, the cost to shift from $t_i^n \in {\cal F}$ to $t_i^{'m} \in {\cal F}'$ being set to $|t_i^n - t_i^{'m}| / \tau$ for a time constant $\tau$.

For small $\tau$, the distance approaches the number of non-coincident spikes, since instead of shifting spikes it is cheaper to insert/delete non-coincident spikes, the distance being always bounded by the number of spikes in both trains.

For high $\tau$ the distance basically equals the difference in spike number (rate distance), while for two spikes trains with the same number of spikes, there is always a time-constant $\tau$ small enough such that the distance is equal to $\sum_n |t_i^n - t_i^{'n}| / \tau$.

Here, two spikes times are comparable if they occur within an interval of $2\,\tau$, otherwise they better are deleted/inserted.

Although computing such distance seems subject to a combinatorial complexity, it appears that quadratic algorithms are available (i.e. with a complexity equal to the product of the number of spikes). {\small This is due to the fact that, in a minimal path, each spike can be either deleted or shifted once to coincide with a spike in the other spike train. Also, a spike can be inserted only at a time that matches the occurrence of a spike in the other spike train.} It allows to calculate iteratively the minimal distance considering the distance $d_{n,n'}({\cal F}, {\cal F}')$ between a spike train composed of the first $n$ spikes of ${\cal F}$ and the first $n'$ spikes of ${\cal F}'$.

%\footnote{A minimal path cannot include an insertion of a spike that is later deleted or shifted, or a deletion of a spike that is later inserted or shifted, or a shift in both direction, since the cost of such path can be reduced by eliminating some steps. Individual spikes cannot intersect, since uncrossing them reduces the amount of shifting.} 

\begin{figure}[htb]  
 \begin{center} \includegraphics[width=7cm,height=1.4cm]{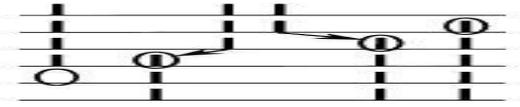} \end{center}
\caption{\small \label{f-to-f} An example of minimal alignment from the upper to the lower spike train, using from top to bottom an insertion, a rightward shift, a leftward shift and a deletion respectively.}
\end{figure}

\vspace{-0.6cm}

When considering spike trains with more than one unit, one point of view is to sum the distances for each alignment unit-to-unit. Another point of view is to consider that a spike can ``jump'', with some cost, from one unit in ${\cal F}$ to another unit in ${\cal F}'$. The related algorithmic complexity is no more quadratic but on the power of the number of units \cite{aronov:03}.

This family of metrics include aligment not only on spike times, but also on inter-spike intervals, or metrics sensitive to motifs of spikes, etc.. They have been fruitfully applied in a variety of neural systems, to characterize neuronal variability and coding \cite{victor:05}. For instance, neurons that act as a coincidence detector with integration time (or temporal resolution) $\tau$, spike trains will have similar postsynaptic effects if they are similar for this metric.

\paragraph{A generalized alignment metric.}

Let us remark, here, that the previous metric generalizes to metric whith:
\\ - [causality] At a given time the cost of previous spikes alignment decreases with the obsolescence of the spike, say, with an exponential time-constant $\tau'$.
\\ - [non-linearity] The cost of a shift can be defined very small, say quadratic, when lower that the time precision and then, say, linear with the time difference.

This leads to an iterative definition of the distance $d_{n,n'}$ defined previously: $d_{n,n'} =$
~\\
\centerline{\small $\mbox{min}\left(\begin{array}{l} e^{-\frac{t_i^n - t_i^{n-1}}{\tau'}} \, d_{n-1,n'} + 1, \\ e^{-\frac{t_i^{'n} - t_i^{'n-1}}{\tau'}} \, d_{n,n'-1} + 1, \\ e^{-\frac{(t_i^n \max t_i^{'n}) - (t_i^{n-1} \min t_i^{'n-1})}{\tau'}} \, d_{n-1,n'-1} + \phi\left(\frac{|t_i^n - t_i^{'n}|}{\tau}\right)\end{array}\right)$}
with, e.g., $\phi(d) = ((d\,\tau/\delta t)^2 \min d$, again implementable in quadratic time. 
It corresponds to the original alignment metric iff $\phi()$ is the identity and $\tau' = +\infty$.

This modified metric illustrates how versatile is this class of distance to represent the differences between spike trains, considering temporal properties only.

\paragraph{Spike training.} As a formal application, let us consider a neuron spike response model \cite{gerstner-kistler:02} of the form:
~\\
\centerline{\small $V_i(t) = \nu(t - t_i^{n-1}) + \sum_{jm} w_{ij} \, \alpha(t - t_j^m)$, $t_i^{n-1} < t \leq t_i^n$,}
the spike time being defined by $V_i(t_i^n) = \theta$, where $\theta$ is the spiking threshold, thus as an implicit equation. 

Previous metrics on spike times allows to optimize the neural weights in order to tune spike-times, deriving, e.g., rules of the form:
~\\
\centerline{$ \Delta w_{ij} \equiv \sum_n (t_i^n - \bar{t}_i^n) \, \frac{\partial V_i}{\partial w_{ij}}(t_i^n) \left/ \frac{\partial V_i}{\partial t_i^n}(t_i^n) \right.$}
where $\bar{\cal F} = \{\cdots, \bar{t}_i^n, \cdots\}$ is the desired spike train

Such mechanisms of optimization is also applicable to time-constants, delays or thresholds. Using spike train metrics open the door to the formalization of such adaptation rules, in order to ``compute with spikes''.

\section{Discussion}

Although probabilistic measures of spike patterns such as correlations \cite{gerstner-kistler:02} or entropy based pseudo-distances (e.g. mutual information) provides a view of spike trains variability which is enriched by the information theory conceptual framework, it may be difficult to estimate them in practice, since such measures are robust only if a large amount of samples is available. On the contrary, deterministic distances allow to characterize aspects of spike coding, with efficient methods and without this curse of sampling size.

This review highlights some of these methods and propose to consider that ``choosing a coding'' means ``defining a metric'', in the deterministic case. 
Making explicit that spikes do not contain that much information.

~\\

{\small {\bf Acknowledgment:} Partially supported by the ANR MAPS \& the EC IP FP6-015879 FACETS projects.}

\bibliographystyle{unsrt0} {\scriptsize \bibliography{string,odyssee,biblio}}

\end{document}